# On-chip arrayed waveguide grating fabricated on thin film lithium niobate

*Zhe Wang,[1] Zhiwei Fang,[1,*] Zhaoxiang Liu,[1] Youting Liang,[1] Jian Liu,[1] Jianping Yu,[1] Ting Huang,[1] Yuan Zhou,[2,3] Haisu Zhang,[1] Min Wang,[1,*] And Ya Cheng[1,2,4,5,6,7,*]*

Corresponding author: zwfang@phy.ecnu.edu.cn; mwang@phy.ecnu.edu.cn; ya.cheng@siom.ac.cn

[1]The Extreme Optoelectromechanics Laboratory (XXL), School of Physics and Electronic Science, East China Normal University, Shanghai 200241, China
[2]State Key Laboratory of High Field Laser Physics and CAS Center for Excellence in Ultra-intense Laser Science, Shanghai Institute of Optics and Fine Mechanics (SIOM), Chinese Academy of Sciences (CAS), Shanghai 201800, China
[3]Center of Materials Science and Optoelectronics Engineering, University of Chinese Academy of Sciences, Beijing 100049, China
[4]State Key Laboratory of Precision Spectroscopy, East China Normal University, Shanghai 200062, China
[5]Collaborative Innovation Center of Extreme Optics, Shanxi University, Taiyuan 030006, China
[6]Collaborative Innovation Center of Light Manipulations and Applications, Shandong Normal University, Jinan 250358, China
[7]Hefei National Laboratory, Hefei 230088, China

**Abstract:** We design an on-chip 8-channel TFLN AWG and fabricate the device using photolithography assisted chemo-mechanical etching (PLACE) technique. We experimentally measure the transmission of the fabricated TFLN AWG near the central wavelength of 1550 nm. We obtain an on-chip loss as low as 3.32 dB, a single-channel bandwidth of 1.6 nm and a total-channel bandwidth of 12.8 nm. The crosstalk between adjacent channels was measured to be below -7.01 dB within the wavelength range from 1543 nm to 1558 nm, and the crosstalk between non-adjacent channels was below -15 dB.

1．Introduction

Monolithically integrated arrayed waveguide grating (AWG) is a photonic integrated circuit device that can combine multiple optical signals of different wavelengths into one waveguide channel or vice versa. In the optical communications industry, AWGs are used as wavelength division multiplexed (WDM) data transmitters for expanding the capacity and improving the transmission rate of optical networks [1-3]. In addition, AWG can also be used in micro-spectrometer and on-chip optical coherence tomography due to their low insertion loss, high spectral resolution, low cost, small footprints and monolithic integration [4,5]. Therefore, significant endeavors have been undertaken to create high-performance AWGs utilizing diverse materials such as silica, silicon, silicon nitride, and polymers [6-12]. Silica-based AWG has been successfully commercialized due to its ultra-low absorption loss (~0.1 dB/km) and mature processing technique, whereas its large footprint (cm$^2$) caused by the low refractive index of silica (~1.44@1550 nm) impedes high-density integration [13]. In contrast, silicon has much higher refractive index (~3.48@1550 nm), enabling significant reduction of the footprint of AWG. Unfortunately, silicon has high absorption loss (~0.5 dB/cm) in the infrared communication band and it cannot operate in visible-band [14-16]. Silicon nitride has a moderately high refractive index (~1.98@1550 nm) as well as low absorption loss (~0.0013 dB/cm) [17,18]. For applications requiring high-speed reconfiguration, the absence of electro-optic effect inhibits silicon-nitride based AWG from fast tuning of its operation wavelength. Polymer-based AWG has the advantages of low cost, flexibility, and ease of fabrication, but it typically has relatively low thermal stability and damage threshold [11,12,19].

Here, we demonstrate an AWG fabricated on the thin film lithium niobate (TFLN) platform. Featured with the high electro-optic coefficient ($r_{33} = 30.9$ pm/V), wide transmission window (0.35-5 μm), low optical loss (~0.002 dB/cm), high refractive index (~2.2), and high thermal stability, and benefitted from the advances of wafer processing technologies, the TFLN becomes an ideal PIC platform for a wide range of applications, such as optical communication, quantum technology, and optical computation [20-29]. We design a low-loss 8-channel AWG on TFLN and fabricate the device using the photolithography assisted chemo-mechanical etching (PLACE) technique. Compared with conventional lithography and reactive ion etching (RIE) process, the PLACE technique not only has large exposure field and high writing speed but also has ultra-low scattering loss [21]. The fabricated TFLN AWG has an on-chip insertion loss as low as 3.32 dB, whereas the insertion loss of the 8-channel TFLN AWG fabricated by lithography and RIE process is 25 dB [30]. The single-channel bandwidth and a total-channel bandwidth are measured as 1.6 nm and 12.8 nm, respectively. The crosstalk between adjacent channels was measured to be less than -7.01 dB within the wavelength range of 1543 nm-1558 nm. Besides, the crosstalk between non-adjacent channels was measured below -15 dB. The fabricated TFLN AWG holds great promise for various applications demanding on-chip wavelength multiplexing/demultiplexing, such as spectral routing of PIC based supercontinuum source and integrated pulse shaper for ultrashort pulses due to fast and efficient electro-optic tuning. The TFLN AWG can also be combined with other optical devices, such as high-speed optical modulators, micro-lasers and waveguide amplifiers, to further improve the performance and application range of TFLN photonic chips [31-36].

**2. Device Design and Fabrication**

In our design, the AWG is to be fabricated on 300-nm thick Z-cut TFLN platform. This not only avoids the generation of higher-order modes in our current waveguide configuration but also maintains the invarient refractive index regardless of the light propagation directions in the plane of TE polarization. Figure 1(a) illustrates the typical cross-section of TFLN rib waveguide fabricated by PLACE technique, which was covered with 1-μm-thick $SiO_2$ film. Figure 1(b) shows the simulated TE mode at 1550 nm in the fabricated TFLN waveguide where the top width is 1.1 μm, the bottom width is 5.2 μm, the etching depth by the chemo-mechanical polish is 250 nm, and the waveguide was supported by the bottom TFLN slab with a thickness of 50 nm.

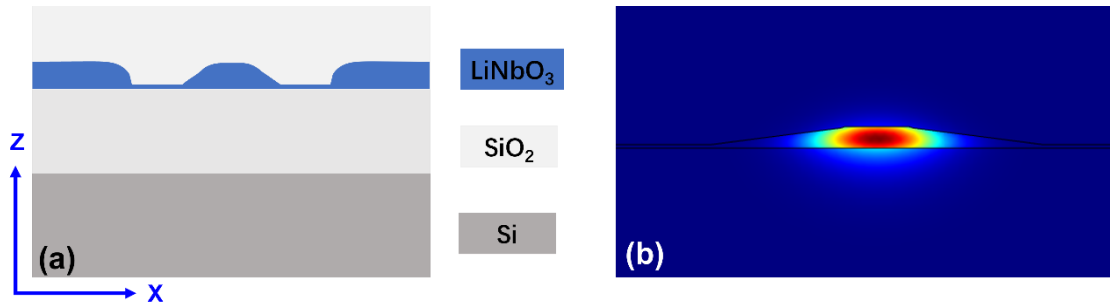

Figure 1. (a) A schematic diagram of cross-section of the fabricated TFLN waveguide. (b) Simulated TE mode profile at λ=1550 nm in z-cut TFLN rib waveguide.

Figure 2(a) is a conceptual illustration of the designed 8-chanels TFLN AWG based on the above waveguide configuration. It is composed of input/output waveguides, two free-propagation regions (FRP) and an array of 20 waveguides. The input/output waveguides are distributed on the circumference of the Roland circle of the two FPRs, and the ends of the array waveguide are located on another side of the FPR, with a fixed length difference ΔL between the adjacent array waveguides. The principle of the AWG is that when a light signal containing multiple wavelengths enters the first FPR from the input waveguide, it diffracts and couples into the array waveguides, and the array waveguides introduce a certain phase shift to the transmitted light signal. For the same wavelength, the output light has the same phase difference, so the light output from the array waveguide interferes in the second FPR, the light of different wavelengths focus on different output waveguide ports and exit from different output waveguides, thus realizing demultiplexing. To ensure the low bending loss of the array waveguide, the overall bending radius is set to be greater than 600 μm. Under the central wavelength of 1550 nm, the effective refractive index of the waveguide is calculated as 1.75158. Based on the working principle of the AWG, we designed the 8-channel AWG at the central wavelength of 1550 nm. In our design, the length of FPR is 610.589 μm, the length difference (ΔL) between adjacent array waveguides is 85.508 μm, and the wavelength spacing of 1.6 nm between adjacent channels. The loss of different channels was simulated and calculated using the beam propagation method as shown in Figure 2(b). According to the simulation results, the loss of different channels was calculated between -2.8 dB and -5.4 dB, and the operating wavelength range of the

AWG device was 12.8 nm with crosstalk between adjacent channels below -10 dB.

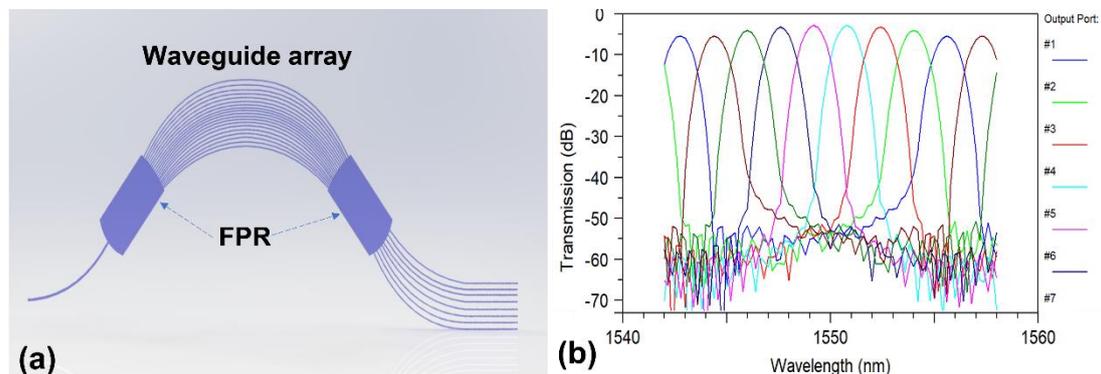

Figure 2. (a) Conceptual illustration of an arrayed waveguide grating. (b) Simulated output spectrum of the 8-channel TFLN AWG around 1550 nm.

The prepared TFLN on insulator wafer is 300-nm thick Z-cut TFLN on 4.7-μm-thick $SiO_2$ film on 500-μm-thick silicon substrate (NANOLN, Jinan Jingzheng Electronics Co. Ltd.), the thickness error of the TFLN is within 10 nm. The TFLN AWG was fabricated by the PLACE technique, the specific processing steps are as follows. Firstly, a 200-nm-thick chromium film was deposited on top of the TFLN using magnetron sputtering. Second, the chromium hard mask pattern for the AWG structure was fabricated by femtosecond laser direct-write lithography. Thirdly, the smooth waveguide edges were etched by chemo-mechanical polish and greatly reduce the scattering loss of waveguide. Finally, to meet the refractive index conditions for single-mode transmission and reduce absorption loss, 1-μm-thick cladding $SiO_2$ film was deposited using inductively coupled plasma chemical vapor deposition (ICPCVD) at low temperature of 80 °C. More fabrication details of the PLACE technique can be found in our previous work [35,37]. Figure 3(a) is a micrograph of the fabricated 8-channel TFLN AWG with a length of 1.7 cm and a width of 0.7 cm. Figure 3(b) is the zoom-in micrograph of the FPR region, where the diffracted light field in the Roland circle is connected to the array waveguide through a taper structure, thereby reducing the insertion loss of the device. Figure 3(c) shows the micrograph of the array waveguide at a magnification factor of 100×, the width of waveguide is 1.1 μm. After chemical mechanical polishing, the edge of the waveguide is extremely smooth, which can greatly reduce the propagation loss of the device.

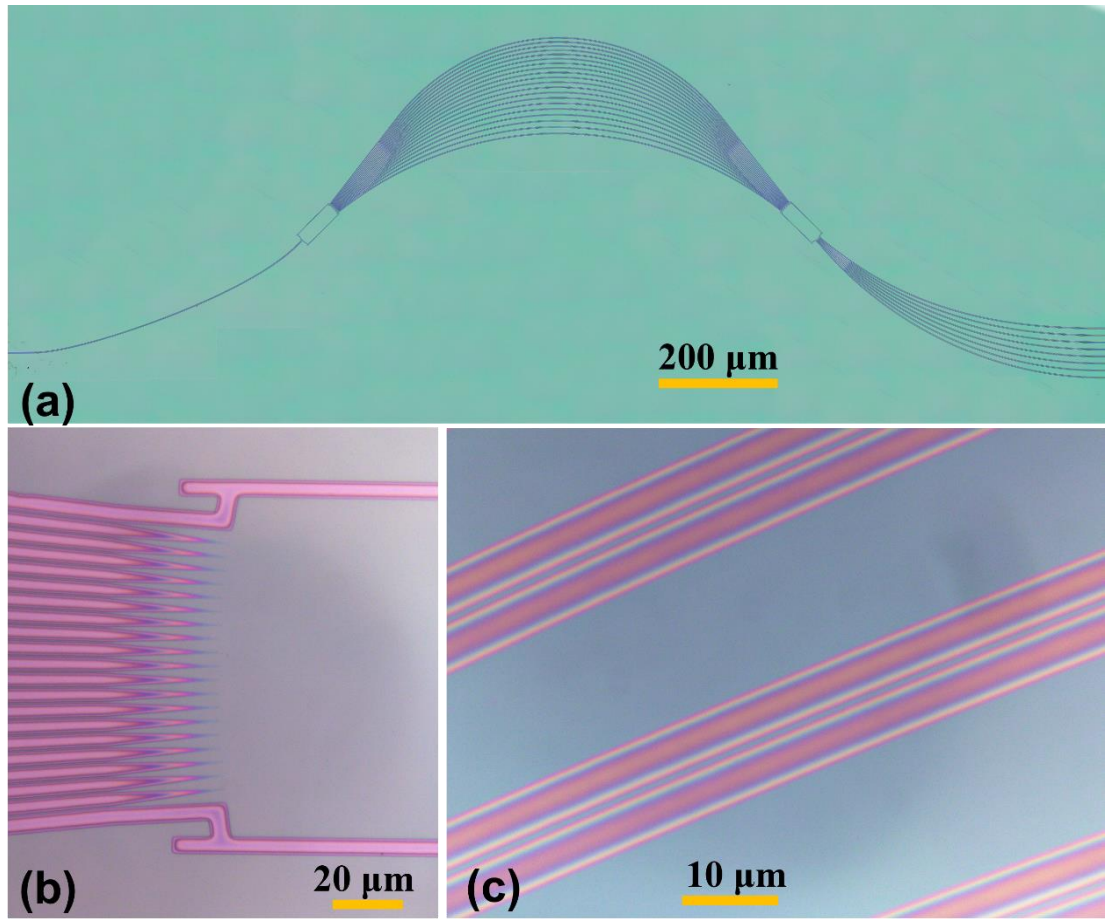

Figure 3. (a)The micrographs of the fabricated 8-channel TFLN AWG. (b)The magnified micrograph of the FPR. (c)The micrograph of the array waveguides.

## 3. Results and Discussions

The optical spectrum measurement device for the TFLN AWG is shown in in Figure 4. A C-band continuously tunable laser (CTL 1550, TOPTICA Photonics Inc.) was coupled to the TFLN AWG using a lensed fiber. The polarization state of both the signal laser are adjusted using an in-line fiber polarization controller (FPC561, Thorlabs Inc.). The output signal from the TFLN AWG is measured by an electrical signal by photodetector (New Focus 1811-FC-AC, Newport Inc.) through lensed fiber. The captured signal is then converted to electrical signals and analyzed by an oscilloscope (Tektronix MDO3104). The couple loss between the lensed fiber and AWG was calculated to be 8.67 dB. When calculating the on-chip insertion loss of the TFLN AWG, the coupling loss was removed.

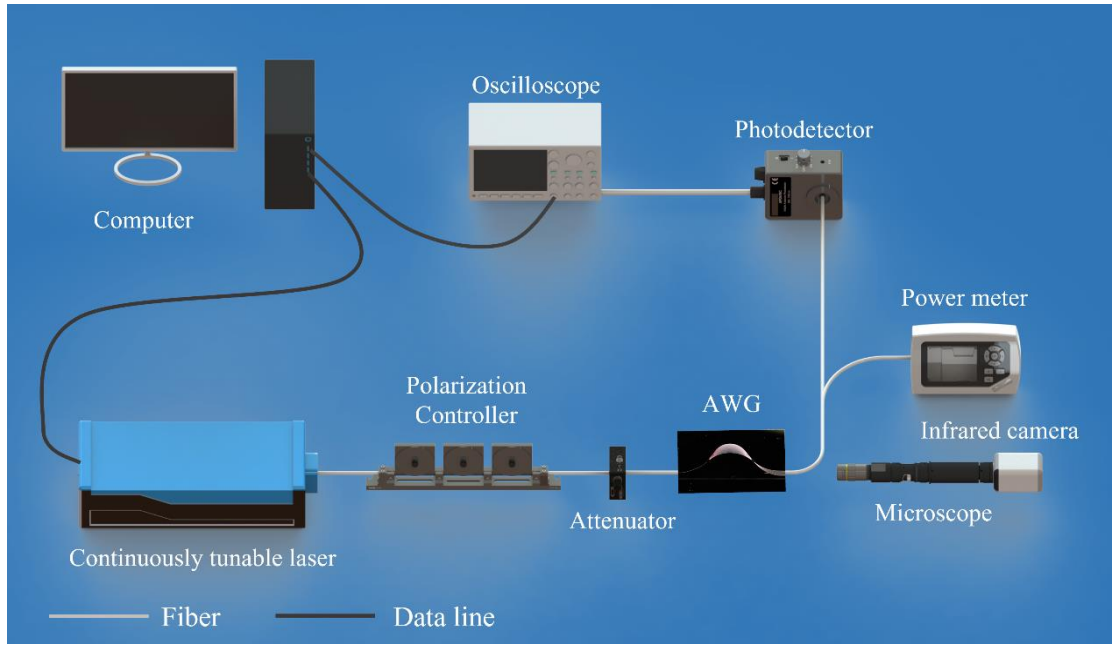

Figure 4. Schematic diagram of optical spectrum measurement setup for the AWG.

As shown in Figure 5(a), the single-channel bandwidth is about 1.6 nm, and a total-channel bandwidth are measured as 12.8 nm. the crosstalk between the adjacent channels is less than -7.1 dB, and the crosstalk between non-adjacent channels is less than -15 dB. The lowest channel loss of the AWG was calculated to be 3.32 dB at a center wavelength of 1552.6 nm in Figure 5(b). In comparison with the simulation results, the insertion loss of the device agrees with the simulation results at the center wavelength, indicating that the waveguide structure on the surface of the prepared device is smooth and has low insertion loss. The reason for the large crosstalk of the device may be related to the adjacent taper area in the input and output regions. The crosstalk maybe reduced by increasing the spacing between tapers at the entrance of waveguide while a trade-off on the insertion loss must be taken into account.

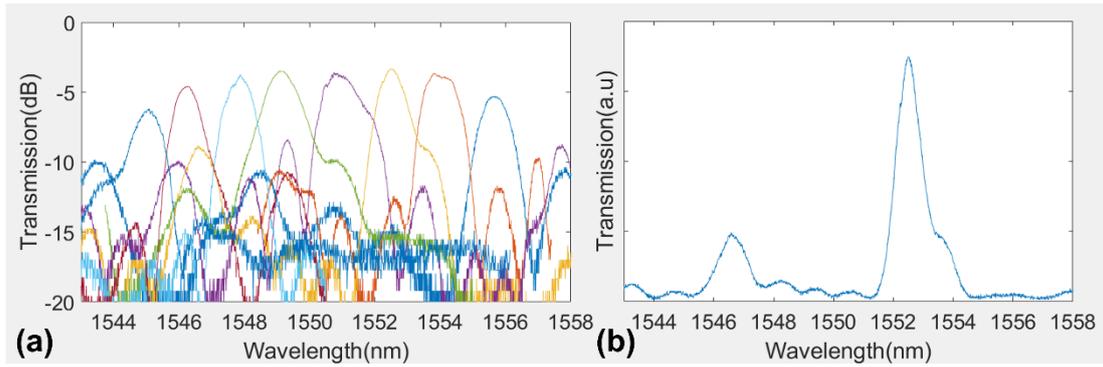

Figure 5. Measurement of the spectrum of the 8-channel AWG. (a) overall spectrum measurement results of the 8 channels. (b) single channel spectrum measurement results.

In our experimental measurement setup, the output of the AWG device can also be imaged by a microscope, and an infrared camera (InGaAs Camera C12741-03, Hamamatsu Photonics Co., Ltd.) is used for observe the output of the AWG when the lensed fiber is removed. As shown in the right panel of Figure 6, the strong optical signals are captured by the infrared camera from their corresponding waveguide channels when a tunable laser is sent into the AWG. The wavelengths measured for the out coming beams are 1555.8 nm, 1554.2 nm, 1552.6 nm, 1550.7 nm, 1549.1 nm, 1547.6 nm, 1546.3 nm, 1544.8 nm from top to bottom, respectively. And the deviation of the central wavelength is within 0.3 nm as compared with the theoretical simulation.

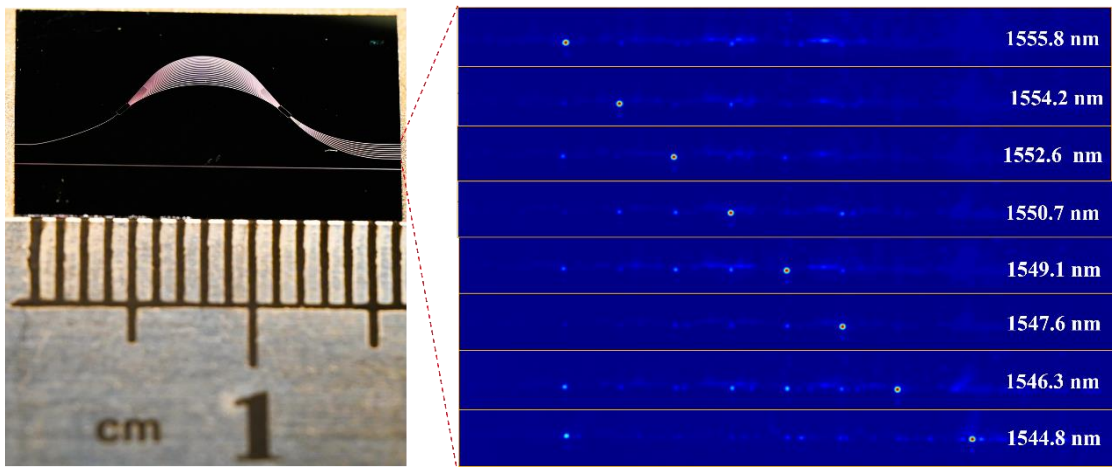

Figure 6. The TFLN AWG chip and the output images captured by the infrared camera when different center wavelengths are injected in the input.

## 4. Conclusion

We successfully fabricate an 8-channel AWG structure on a 300 nm Z-cut TFLN using PLACE technique. The measured insertion losses of the device are in the range of 3.1 dB-6.2 dB. The measured spectral bandwidth of the AWG is 12.8 nm. The AWG device can be applied in the field of high-capacity communication, and integratable with other on-chip TFLN devices such as on-chip lasers and detectors to achieve miniaturized multi-functional photonic chips.

**Acknowledgements**

National Key R&D Program of China (2019YFA0705000, 2022YFA1404600, 2022YFA1205100), National Natural Science Foundation of China (Grant Nos. 12004116, 12274133, 12104159, 11874154, 11734009, 11933005, 12134001, 61991444), Science and Technology Commission of Shanghai Municipality (NO.21DZ1101500), Shanghai Municipal Science and Technology Major Project (Grant No.2019SHZDZX01), Shanghai Sailing Program (21YF1410400) and Shanghai Pujiang Program (21PJ1403300). Innovation Program for Quantum Science and Technology (2021ZD0301403).

Received: ((will be filled in by the editorial staff))
Revised: ((will be filled in by the editorial staff))
Published online: ((will be filled in by the editorial staff))

**References**

[1] Q. Cheng, M. Bahadori, M. Glick, S. Rumley, and K. Bergman, "Recent advances in optical technologies for data centers: a review," Optica 5, 1354-1370 (2018)

[2] M. K. Smit and C. Van Dam, "PHASAR-based WDM-devices: Principles, design and applications," IEEE J. Sel. Top. Quantum Electron. 2, 236-250 (1996).

[3] T. Tsuchizawa, K. Yamada, T. Watanabe, S. Park, H. Nishi, K. Rai, H. Shinojima, and S.-I. Itabashi, "Monolithic integration of silicon-, germanium-, and silica-based optical devices for telecommunications applications," IEEE J. Select. Top.Quantum Electron 17(3), 516–525 (2011).

[4] P. Cheben, J. H. Schmid, A. Delâge, A. Densmore, S. Janz, B. Lamontagne, J.

Lapointe, E. Post, P. Waldron, and D.-X. Xu, "A high-resolution silicon-on-insulator arrayed waveguide grating microspectrometer with sub-micrometer aperture waveguides," Opt. Express 15, 2299-2306 (2007)

[5] E. A. Rank, R. Sentosa, D. J. Harper, M. Salas, A. Gaugutz, D. Seyringer, S. Nevlacsil, A. Maese-Novo, M. Eggeling, P. Muellner, R. Hainberger, M. Sagmeister, J. Kraft, R. A. Leitgeb, and W. Drexler, "Toward optical coherence tomography on a chip: in vivo three-dimensional human retinal imaging using photonic integrated circuit-based arrayed waveguide gratings," Light: Sci. Appl. 10(1), 6 (2021).

[6] Y. Hibino, "Recent advances in high-density and large-scale AWG multi/demultiplexers with higher index-contrast silica-based PLCs," IEEE J. Sel. Top. Quantum Electron. 8, 1090-1101 (2002).

[7] S. Kamei, Y. Doi, Y. Hida, Y. Inoue, S. Suzuki, and K. Okamoto, "Low-loss and flat/wide-passband CWDM demultiplexer using silica-based AWG with multi-mode output waveguides," in Optical Fiber Communication Conference, Technical Digest (CD) (Optica Publishing Group, 2004), paper TuI2.

[8] D. Dai, and J. E. Bowers, "Silicon-based on-chip multiplexing technologies and devices for Peta-bit optical interconnects" Nanophotonics, 3, 283-311(2014).

[9] D. Dai, Z. Wang, J. F. Bauters, M. C. Tien, M. J. R. Heck, D. J. Blumenthal, and J. E. Bowers, "Low-loss $Si_3N_4$ arrayed-waveguide grating (de)multiplexer using nano-core optical waveguides," Opt. Express 19, 14130– 14136 (2011).

[10] D. Martens, S. Selvaraja, P. Verheyen, G. Lepage, W. Bogaerts, and D. Van Thourhout, "Compact silicon nitride arrayed waveguide gratings for very near-infrared wavelengths," IEEE Photon. Technol. Lett. 27, 137-140 (2015).

[11] H. Ma, A.K. -Y. Jen, and L.R. Dalton, "Polymer-Based Optical Waveguides: Materials, Processing, and Devices," Adv. Mater., 14: 1339-1365(2002).

[12] B. Yang, Y. Zhu, Y. Jiao, L. Yang, Z. Sheng, and D. Dai, "Compact arrayed waveguide grating devices based on small SU-8 strip waveguides," J. Lightwave Technol. 29, 2009-2014(2011).

[13] A. Stoll, K. Madhav, and M. Roth, "Performance limits of astronomical arrayed waveguide gratings on a silica platform," Opt. Express 28, 39354-39367 (2020).

[14] H. H. Li, "Refractive index of silicon and germanium and its wavelength and temperature derivatives," J. Phys. Chem. Ref. Data 9, 561–658 (1980).

[15] R. Huang, Y. Zhao, X. She, H. Liao, J. Zhu, Z. Zhu, X. Liu, H. Liu, Z. Sheng, and F. Gan, "High resolution, high channel count silicon arrayed waveguide grating router on-chip," Opt. Express 31, 14308-14316 (2023).

[16] S. K. Selvaraja, P. D. Heyn, G. Winroth, P. Ong, G. Lepage, C. Cailler, A. Rigny, K. K. Bourdelle, W. Bogaerts, D. V. Thourhout, J. V. Campenhout, and P. Absil, "Highly uniform and low-loss passive silicon photonics devices using a 300 mm CMOS platform," in Optical Fiber Communication Conference (2014), paper Th2A.33.

[17] A. Arbabi and L. L. Goddard, "Measurements of the refractive indices and thermo-optic coefficients of $Si_3N_4$ and $SiO_x$ using microring resonances," Opt. Lett. 38, 3878–3881(2013).

[18] X. Ji, F. A. S. Barbosa, S. P. Roberts, A. Dutt, J. Cardenas, Y. Okawachi, A. Bryant, A. L. Gaeta, and M. Lipson, "Ultra-lowloss on-chip resonators with sub-milliwatt parametric oscillation threshold," Optica 4, 619–624 (2017).

[19] Y. Yu, Z. Yu, Z. Zhang, H. K. Tsang, and X. Sun, "Wavelength-division multiplexing on an etchless lithium niobate integrated platform," ACS Photonics 9, 3253–3259 (2022).

[20] D. N. Nikogosyan, Nonlinear Optical Crystals: A Complete Survey (Springer Science & Business Media, 2006).

[21] R. Gao, N. Yao, J. Guan, L. Deng, J. Lin, M. Wang, L. Qiao, W. Fang, and Y. Cheng, "Lithium niobate microring with ultra-high Q factor above $10^8$," Chin. Opt. Lett. 20, 011902(2022).

[22] A. Shams-Ansari, G. Huang, L. He, Z. Li, J. Holzgrafe, M. Jankowski, M. Churaev, P. Kharel, R. Cheng, and D. Zhu, N. Sinclair, B. Desiatov, M. Zhang, T. J. Kippenberg, and M. Lončar, "Reduced material loss in thin-film lithium niobate waveguides," APL Photon. 7, 081301 (2022).

[23] A. Boes, B. Corcoran, L. Chang, J. Bowers, and A. Mitchell, "Status and Potential of Lithium Niobate on Insulator (LNOI) for Photonic Integrated Circuits," Laser

Photon. Rev. 12, 1700256 (2018).

[24] A. Honardoost, K. Abdelsalam, S.Fathpour, "Rejuvenating a Versatile Photonic Material: Thin-Film Lithium Niobate," Laser Photon. Rev. 14, 2000088(2020).

[25] J. Lin, F. Bo, Y. Cheng, and J. Xu, "Advances in on-chip photonic devices based on lithium niobate on insulator," Photon. Res. 8, 1910 (2020).

[26] Y. Jia, L. Wang, and F. Chen, "Ion-cut lithium niobate on insulator technology: Recent advances and perspectives", Appl. Phys. Rev. 8, 011307 (2021).

[27] S. Saravi, T. Pertsch, F. Setzpfandt, "Lithium niobate on insulator: an emerging platform for integrated quantum photonics", Adv. Opt. Mater. 9, 2100789(2021).

[28] G. Chen, N. Li, J. D. Ng, H. -L. Lin, Y. Zhou, Y. H. Fu, L. Y. T. Lee, Y. Yu, A. -Q. Liu, A. J. Danner, "Advances in lithium niobate photonics: development status and perspectives," Adv. Photon. 4 034003 (2022).

[29] Y. Zheng, H. Zhong, H. Zhang, R. Wu, J. Liu, Y. Liang, L. Song, Z. Liu, J. Chen, J. Zhou, Z. Fang, M. Wang, and Ya Cheng "Electro-optically programmable photonic circuits enabled by wafer-scale integration on thin-film lithium niobate." arXiv preprint arXiv:2304.03461 (2023).

[30] M. Prost, G. Liu, and S. J. Ben Yoo, "A compact thin-film lithium niobate platform with arrayed waveguide gratings and MMIs," in Optical Fiber Communication Conference, OSA Technical Digest (online) 2018, paper Tu3A.3.

[31] C. Wang, M. Zhang, X. Chen, M. Bertrand, A. Shams-Ansari, S. Chandrasekhar, P. Winzer, and M. Loncar, "Integrated lithium niobate electro-optic modulators operating at CMOS-compatible voltages," Nature 562, 101–104 (2018).

[32] M. Xu, M. He, H. Zhang, J. Jian, Y. Pan, X. Liu, L. Chen, X. Meng, H. Chen, Z. Li, X. Xiao, S. Yu, S. Yu, and X. Cai, "High-performance coherent optical modulators based on thin-film lithium niobate platform," Nat. Commun. 11, 3911 (2020).

[33] R. Wu, L. Gao, Y. Liang, Y. Zheng, J. Zhou, H. Qi, D. Yin, M. Wang, Z. Fang, and Y. Cheng, "High-Production-Rate Fabrication of Low-Loss Lithium Niobate Electro-Optic Modulators Using Photolithography Assisted Chemo-Mechanical Etching (PLACE)," Micromachines 13, 378 (2022).

[34] Y. Han, X. Zhang, F. Huang, X. Liu, M. Xu, Z. Lin, M. He, S. Yu, R. Wang, and X. Cai, "Electrically pumped widely tunable O-band hybrid lithium niobite/III-V laser," Opt. Lett. 46, 5413-5416 (2021)

[35] Y. Liang, J. Zhou, R. Wu, Z. Fang, Z. Liu, S. Yu, D. Yin, H. Zhang, Y. Zhou, J. Liu, Z. Wang, M. Wang, and Y. Cheng, "Monolithic single-frequency microring laser on an erbium-doped thin film lithium niobate fabricated by a photolithography assisted chemo-mechanical etching," Opt. Continuum 1, 1193-1201 (2022).

[36] J. Zhou, Y. Liang, Z. Liu, W. Chu, H. Zhang, D. Yin, Z. Fang, R. Wu, J. Zhang, W. Chen, Z. Wang, Y. Zhou, M. Wang, and Y. Cheng, "On-chip integrated waveguide amplifiers on Erbium-doped thin film lithium niobate on insulator," Laser Photon. Rev. 15, 2100030(2021).

[37] M. Wang, R, Wu, J. Lin, J, Zhang, Z. Fang, Z, Chai and Y. Cheng, "Chemo-mechanical polish lithography: A pathway to low loss large-scale photonic integration on lithium niobate on insulator," Quantum Engineering 1, e9 (2019).